\documentclass{article}
\usepackage{epic,eepic,psfig}
\usepackage{amssymb}
\usepackage{code}

\newcounter{cnt}
\newenvironment{example}{
        \def\thecnt{\arabic{cnt}}
        \refstepcounter{cnt}
        \trivlist\item[\hskip
        \labelsep{\bf Example~\thecnt.}]}{\framebox[2.1mm]{~}\endtrivlist}

\newcommand{\ignore}[1]{}

\newcommand{\martinscomment}[1]{}
\newcommand{\kevinscomment}[1]{}
\newcommand{\peterscomment}[1]{}
\newcommand{\pjs}[1]{}
\newcommand{\kg}[1]{}

\newcommand{\simparrow}[0]{\Longleftrightarrow}        
\newcommand{\proparrow}[0]{\Longrightarrow}        
        
\newcommand{\arrow}[0]{\mapsto}

\newcommand{\kindone}[0]{\mbox{\tt Kind*->*}}
\newcommand{\kindzero}[0]{\mbox{\tt Kind*}}

\newcommand{\mclass}[0]{\mbox{\it class}}
\newcommand{\mClass}[0]{\mbox{\it C}}
\newcommand{\minst}[0]{\mbox{\it instance}}

\newcommand{\mwhere}[0]{\mbox{\it where}}

\newcommand{\mlabel}[1]{#1}
\newcommand{\atsign}{@}

\newcommand{\ba}{\begin{array}}
\newcommand{\ea}{\end{array}}

\def\tuple#1{\langle #1 \rangle}
\title{\bf Type Classes and Constraint Handling Rules}
\author{Kevin Glynn, Peter J. Stuckey and Martin Sulzmann \\
Dept.\ of Computer Science and Software Engineering, \\
The University of Melbourne, \\
Parkville 3052, Australia. \\
{\tt\{keving,pjs,sulzmann\}@cs.mu.oz.au}
}
\date{}

\begin{document}

\bibliographystyle{alpha}
\maketitle

\makeatactive
\begin{abstract}
Type classes are an elegant extension to traditional, Hindley-Milner
based typing systems. 
They are used in modern, typed languages such as Haskell to support controlled
overloading of symbols. 
Haskell 98 supports only single-parameter and constructor type classes.
Other extensions such as multi-parameter type classes are highly desired 
but are still not officially supported by Haskell.
Subtle issues arise with extensions, which may lead to a loss of feasible type inference
or ambiguous programs.
A proper logical basis for type class systems seems to be missing. Such a basis would
allow extensions to be characterised and studied rigorously.
We propose to employ Constraint Handling Rules as a tool to study and develop
type class systems in a uniform way.
\end{abstract}

\section{Introduction}

The Haskell language~\cite{haskell98} provides one of the most advanced type
systems in an industrial--strength 
language. Type classes are one of the most distinctive features of Haskell.
The form of type classes found in Haskell 98 is restricted to single-parameter and 
constructor~\cite{jones:overloading} type classes.
A rigorous treatment of the Haskell 98 type system can be found in 
in~\cite{jones:typing-haskell-in-haskell}, it fills some serious gaps in the current specifications of Haskell.

Since the original papers~\cite{kaes:parametric,wadler-blott:ad-hoc} on type classes,
many researchers have studied extensions to the existing type class system~\cite{para-type-class,jones-jones-meijer:typeclasses}.
In particular, multi-parameter type classes are a very desirable extension,
see~\cite{jones-jones-meijer:typeclasses} for an overview.
However, multi-parameter type classes are still not officially supported by Haskell.
As the authors note, design decisions need to be taken with
great care in order to retain feasible type inference.

Existing implementations of Haskell use a dictionary passing translation to support type class overloading.
This requires types to be unambiguous, otherwise an implementation can't know which 
dictionary should be passed.
Unfortunately, even with the existing single-parameter class system, ambiguous types can occur. 
In his recent paper~\cite{JonesESOP2000}, Jones extends type classes with functional dependencies
to resolve ambiguity in the context of multi-parameter classes for certain cases.
We find that type classes are still an active area of research and the debate about which
features should be incorporated into future Haskell specifications is far from settled.

\mbox{} \\
{\bf This work.}
In contrast to previous work, our foremost goal is not to propose 
yet another extension of type classes.
The thesis of this paper is:
{\em Constraint handling rules~\cite{fruehwirth:CHRs} 
are the right way to understand
type class constraints, and extensions to type classes.
In particular, constraint handling rules 
help us understand the two main 
issues behind possible type class extensions:
Feasible type inference and unambiguous programs.}

Feasible type inference is an important property which needs to be
retained when considering type class extensions.  Type
inference should be decidable and should compute principal types.
Hindley-Milner types are characterized by constraints that are
representable using Herbrand constraints, and solvable using
well-understood constraint solvers such as Robinson's unification
algorithm.  Clearly, type classes are simply another form of constraint system,
extending the Herbrand constraints, that constrains the value that
various type variables can take.  

Constraint handling rules (CHRs) are a way
of extending constraint solving from a well-understood underlying
constraint domain to handle new forms of constraints.  CHRs 
are a simple language and efficient implementations are available.
They give a natural definition of a constraint solver for type classes, they clarify
some issues about what meaning should be given to type class extensions and they
give insight into problems such as ambiguity, overlapping instances of type
classes and multi-parameter type classes.
Moreover, CHRs allow us to specify the conditions under which feasible type inference is guaranteed.

For the purpose of this workshop paper, we explain our ideas by example rather than
by giving a rigorous formal treatment.

\mbox{} \\
{\bf Outline.}
Section~\ref{sec:motivation} motivates our approach by reviewing limitations with Haskell's 
type class system, which can be overcome with extensions defined by CHRs. 
In Section~\ref{sec:chr}, we review the basic ideas behind CHRs.
Type inference is described in Section~\ref{sec:inference}.
Section~\ref{sec:properties} states some sufficient conditions
under which we achieve feasible type inference and unambiguous programs.
In Section~\ref{sec:extensions}, we show how to express some (previously proposed in the literature)  
type class extensions in terms of CHRs.
In Section~\ref{sec:new-extensions}, we show that 
CHRs prove to be useful for defining novel type class extensions.
We conclude in Section~\ref{sec:conclusion}.

\section{Motivation} \label{sec:motivation}

Type classes are an elegant extension to traditional, Hindley-Milner
based typing systems.  In addition to supporting controlled
overloading of functions they allow programmers (and language
designers) to identify related types.  For example, when we make our
new type an instance of the @Eq@ class; not only are we adding the
convenience of the @==@ operator we are telling users of this type
that its inhabitants are exact and identifiable.  Note that if we
choose not to make our type an instance of @Eq@ we are also making a
statement: inhabitants are representing values which are inexact
and/or hard to identify.

Unfortunately, in Haskell 98 it is impossible to enforce any
restrictions on class membership.  Worse, instance declarations are
global and a program can only have one instance declaration for a type
and a particular class.  This requires that any instance declaration
visible in a module is exported to all modules which import it,
bypassing Haskell's name hiding mechanisms.  So, if a module declares
@Bool@ to be a member of the @Num@ class then any importing module
also treats Booleans as a member of @Num@.  The writer of the
importing module may be completely unaware that this has happened. Now
`errors' such as
\begin{quote}
\begin{code}
import X()                         -- X makes Bool an instance of Integral

cap_power :: (Num a,Integral b) => Bool -> b -> a -> b -> a
cap_power useLimit theLimit n p 
    | useLimit && theLimit < p = n ^ useLimit   -- surely intended `theLimit'
    | otherwise                = n ^ p
\end{code}
\end{quote}
will compile just fine, though with bizarre consequences.

Many researchers have studied extensions which would make them more
useful, while maintaining Haskell's decidable type inference.  In
particular, extensions have been proposed to support multi parameter
type classes and allow relationships amongst components of a type
class other than the standard super/sub class relations.  These
extensions are ad-hoc, requiring special syntax and `hidden'
restrictions to be added to the language in order to guarantee
principal types and a decidable type inference algorithm.

CHRs provide a framework which allows much more freedom for the
programmer and language designer to specify constraints on type
classes.  These extra constraints can serve both to make more programs
typable (since the more powerful constraints remove potential
ambiguity of types in a program) and to make less programs typable
(since the class constraints can restrict the use of types to those
intended by the class writer).

CHRs will be described in the next section,  but for the remainder of
this section we will introduce some examples to give a feel for how CHRs can
make the language both safer and more expressive. 

\subsection{Disjoint Classes}

Haskell 98's support for type classes is too restrictive even without
adding multi-parameter type classes to the language. For example, we
would like to say that the @Integral@ and @Fractional@ type classes
are disjoint.  This cannot be expressed in current Haskell and hence
the function
\begin{quote}
\begin{code}
f x y = x / y + x `div` y
\end{code}
\end{quote}
has an inferred type of 
@f :: (Integral a, Fractional a) => a -> a -> a@
rather than immediately causing a type error.  

Disjoint classes can also be used to resolve ambiguities in a program.
Overlapping type class instances are a thorny issue for Haskell
implementations.  In general, overlapping instances lead to
unacceptable ambiguity in programs.  But with improved class
information it would sometimes be clear that overlapping instances do
not really overlap at all.  For example imagine a ``has division
operator'' class @Dividable@ defined as follows:
\begin{quote}
\begin{code}
class Num a => Dividable a where dividedBy :: a -> a -> a

instance Fractional a => Dividable a where dividedBy = (/)
instance Integral a => Dividable a where dividedBy = div

halfish :: Dividable a => a -> a
halfish x = x `dividedBy` 2
\end{code}
\end{quote}
Although the instances appear to overlap, we know that the @Integral@
and @Fractional@ type classes are intended to be disjoint, therefore
there should be no ambiguity here.  Notice that this also implies a
further extension to GHC's existing, experimental support for
overlapping instances. Currently, GHC ignores the instance constraints
when deciding if two instance declarations overlap.

A special case of disjoint sets allows us to specify instance
declarations for certain types to be an error.  For example, we could
add constraints to the prelude so that programmers can't make types
such as @Bool@, @Char@ and function types instances of the @Num@
class. This can be done directly for each type or we can declare a
disjoint set for @Num@, e.g. @NotNum@, and make the types instances of
this disjoint class.

\subsection{Multi-Parameter Classes}

As Jones~\cite{JonesESOP2000} points out, the need to extend type
classes to a relation over types is well understood, and most Haskell
implementations support multi-parameter type classes by non-standard extensions.  However, in
practice they haven't worked as well as was hoped.  Many useful programs have 
ambiguous types or fall foul of syntactic restrictions required for feasible
type inference. By
allowing more expressive constraints between the elements of a class
relation, these programs can be typed, and in addition the class designer  has finer control over 
their use. Below, we show
some useful relationships using the @Collects@ class, as presented by Jones,  as an example.
\begin{quote}
\begin{code}

class Collects e ce where
    empty :: ce
    insert :: e -> ce -> ce
    member :: e -> ce -> Bool
\end{code}
\end{quote}

\begin{description}
\item[Functional Dependencies] Jones proposes to extend type classes
  with {\em Functional Dependencies} to support dependencies amongst
  Class components.  These give the programmer the necessary control
  over the allowed relationships to allow additional programs to be
  typed.  The @empty@ method of @Collects@ has an ambiguous type (since only @ce@ 
  is fixed and there may be more than one possible instance declaration which could be used). 
  But
  if we say that the type of @e@ is {\it dependent} on the type of @ce@, it
  is no longer ambiguous.  CHRs subsume the expressiveness of
  functional dependencies.  The remaining examples are not possible
  with only functional dependencies, however they are supported by
  CHRs.
  
\item[Anti-symmetrical Relations] The @Collects@ class allows us to
  declare that @[a]@ is a collection
  of @a@s by an instance declaration for @Collects a [a]@. 
  However, it would never be sensible to declare an instance for
  the reverse, i.e. @Collects [a] a@.  CHR's can specify that the relationship is
  anti-symmetrical, making the latter instance declaration a
  type error.
  
\item[Irreflexive Relations] The @Collects@ relation is also
  irreflexive, it wouldn't be sensible to allow a type to be a collection type for itself, 
  i.e. @Collects@ $\tau_1~~\tau_2$ where $\tau_1$ and $\tau_2$ are unifiable. 
 
\end{description}

Many other relationships, such as transitivity or symmetry, can also
be expressed with CHRs.

\subsection{Constructor Classes}

Type constructors, such as List (actually, @[]@) or @->@, are functions over
types. Implicitly each type constructor has a ${\mbox{\it Kind}}$ which specifies the number of
argument types required to produce the result type.

Constructor classes, which are supported by Haskell 98, allow the
programmer to write a class over type constructors.  Then, any type
constructor with the correct kind can be made an instance of the
class,  e.g. the definition of the @Functor@ class in Haskell 98 is:
\begin{quote}
\begin{code}
class Functor f where fmap :: (a -> b) -> f a -> f b
\end{code}
\end{quote}
Instances can be provided for @[]@ and @Tree@ but not for @Int@ or
@->@, since they don't have the correct kind. 

Instead of introducing a kind system to Haskell, we could use CHRs to
express these constraints as multi-parameter class constraints.  The 
CHR system can explicitly describe the required constraints, i.e. the
type constructor is functional, surjective and correctly kinded.

\section{Constraint Handling Rules} \label{sec:chr}

Constraint handling rules~\cite{fruehwirth:CHRs} (CHRs) 
are a multi-headed concurrent constraint language for
writing incremental 
constraint solvers.  In effect, they define transitions from
one constraint set to an equivalent constraint set. Transitions serve to
simplify constraints and detect satisfiability and unsatisfiability.
Efficient implementations of CHRs are available in the languages
SICStus Prolog and Eclipse Prolog, and other implementations are currently 
being developed (e.g.,~for Java).

CHRs manipulate a constraint set in two parts:
a global constraint in the language of the underlying solver, and
a global set of primitive 
constraints defined only by constraint handling rules.

Constraint handling rules (\emph{CHR rules}) 
are of two forms (though the first is sufficient)
\begin{eqnarray*}
{\mbox {\it simplification}} &&
rulename \atsign c_1, \ldots, c_n 
        ~~\simparrow~~  g ~|~ d_1, \ldots, d_m \\
{\mbox {\it propagation}} &&
rulename \atsign c_1, \ldots, c_n 
        ~~\proparrow~~  g ~|~ d_1, \ldots, d_m
\end{eqnarray*}
In these rules $rulename$ is a unique identifier for a rule,
$c_1, \ldots, c_n$ are CHR constraints, $g$
is a Herbrand constraint, and $d_1, \ldots, d_m$ are either
CHR or Herbrand constraints. The guard part $g$ is optional.
When it is omitted it is equivalent to $g \equiv True$. 
The simplification rule states that given a constraint set 
$\{c_1, \ldots, c_n\}$ where $g$ must hold, this set 
can be replaced by $\{d_1, \ldots, d_m\}$.
The propagation rule states that given a constraint set 
$\{c_1, \ldots, c_n\}$ where $g$ must hold, we should add 
$\{d_1, \ldots, d_m\}$.
A \emph{CHR program} is a set of CHR rules.

More formally the logical interpretation of the rules is as follows.
Let $\bar{x}$ be the variables occurring in $\{c_1, \ldots, c_n\}$,
and $\bar{y}$ (resp.~$\bar{z}$) be the other variables occurring in
the guard $g$ (resp.~rhs $d_1, \ldots, d_m$) of the rule. 
We assume no local variables appear in both the guard and the rhs.
The logical reading is
\begin{eqnarray*}
{\mbox {\it simplification}} &&
\forall \bar{x} (\exists \bar{y}~ g) \rightarrow ( c_1 \wedge \cdots \wedge c_n 
        \leftrightarrow (\exists \bar{z}~ d_1 \wedge \cdots \wedge d_m)) \\
{\mbox {\it propagation}} &&
\forall \bar{x} (\exists \bar{y}~ g) \rightarrow ( c_1 \wedge \cdots \wedge c_n 
        \rightarrow (\exists \bar{z}~d_1 \wedge \cdots \wedge d_m)) 
\end{eqnarray*}

The operational semantics (see~\cite{abdennadher:confluence} for more detail) 
is a transition system on a quadruple
$\tuple{f,s,h,t}_v$ of a conjunction of CHR and Herbrand constraints
$f$, a conjunction of CHR constraints $s$, a conjunction of
Herbrand constraints $h$, a set of tokens for use in controlling termination
and a sequence of variables $v$.  
The logical reading of $\tuple{f,s,h,t}_v$ is as 
$\exists \bar{y} f \wedge s \wedge h$ where $y$ are the variables in
the tuple not in $v$.  Since the variable component $v$ never changes
we omit it for much of the presentation.

Tokens take the form $r\atsign{}h$ where $r$ is a rulename of a propagation rule, 
and $h$ is a conjunction of CHR constraints matching the left hand side of
the rule.  In order to avoid trivial non-termination with propagation rules,
which if they are applied once can be applied again, the operational
semantics restricts that a propagation rule can only be applied once to
each set of matching CHR constraints.  Tokens are introduced when a new
CHR constraint is moved to the store $s$, and used up on the application
of a propagation rule. Define $token(c,s)$ to be the set of new 
propagation rule applications which could be applied when adding CHR
constraint $c$ to a CHR store $s$. 
$$
token(c,s) = \{ r\atsign{}c \wedge c' ~|~ (r \atsign{} c'' \proparrow g | b) \in P, s = c' 
\wedge s', \models \exists ((c \wedge c') = c'') \}  
$$
Tokens are removed through the use of propagation rules, and when 
they are no longer applicable through a simplification

\begin{center}
\begin{tabular}{lll}
solve & $\tuple{d \wedge f, s,h, t} \rightarrowtail \tuple{f, s, h', t}$
& \\
& $d$ is a Herbrand constraint, $\models h' \leftrightarrow h \wedge d$ \\
introduce & $\tuple{d \wedge f, s,h, t} \rightarrowtail \tuple{f, s \wedge d,
h, t \cup token(c,s)}$& \\
& $d$ is a CHR constraint \\
simplify & $\tuple{f, c' \wedge s,h, t} \rightarrowtail_P \tuple{d \wedge f, s, 
h \wedge c = c', t - \cup_{c'_i \in c'} token (c'_i,c'\wedge s)}$  \\
& $r \atsign{} c \simparrow g ~|~ d$ in $P$ and 
$\models h \rightarrow \exists \bar{x} (c = c' \wedge g)$ \\
propagate & $\tuple{f, c' \wedge s,h, t \cup \{r\atsign{}c'\}} \rightarrowtail_P
\tuple{d \wedge f, c' \wedge s, h \wedge c = c', t}$ \\
& $r \atsign{} c \proparrow g ~|~ d$ in $P$ and 
$\models h \rightarrow \exists \bar{x} (c = c' \wedge g)$ \\
\end{tabular}
\end{center}
where $\bar{x}$ are variables (assumed to be new) 
appearing in the CHR used.
Note that the components of the triple are treated as 
conjunctions and the matching is modulo the idempotence, commutativity,
and associativity of conjunction.

An important property of CHR programs is \emph{confluence}.
Confluence implies that the order of the transitions doesn't affect the
final result. 
Two states $\tuple{f_1,s_1,h_1,t_1}_v$ and $\tuple{f_2,s_2,h_2,t_2}_v$ are
\emph{joinable} if there exists derivations 
$\tuple{f_1,s_1,h_1,t_1}_v \rightarrowtail^*_P \tuple{f_3,s_3,h_3,t_3}_v$
and 
$\tuple{f_2,s_2,h_2,t_2}_v \rightarrowtail^*_P \tuple{f_4,s_4,h_4,t_3}_v$
such that $\tuple{f_3,s_3,h_3,t_3}_v$ is a variant of 
$\tuple{f_4,s_4,h_4,t_4}_v$.
Confluent CHR programs are guaranteed to be \emph{consistent} (in the usual
sense of a theory).

A CHR program $P$ is confluent iff for each
state $\tuple{f,s,h,t}_v$, if 
$\tuple{f,s,h,t}_v \rightarrowtail^*_P \tuple{f_1,s_1,h_1,t_1}_v$
and $\tuple{f,s,h.t}_v \rightarrowtail^*_P \tuple{f_2,s_2,h_2,t_2}_v$
then $\tuple{f_1,s_1,h_1,t_1}_v$ and $\tuple{f_2,s_2,h_2,t_2}_v$ are joinable.

Importantly, for terminating CHR programs, confluence is
\emph{decidable}~\cite{abdennadher:confluence} 
(although termination is not decidable).  This is because 
for these programs, confluence is equivalent to local confluence
which we can test by examining each \emph{critical pair}
of the program and seeing whether they are joinable. 
A critical pair of two rules
\begin{eqnarray*}
r_1 \atsign{} c_1 \wedge c_1' \simparrow g_1 ~|~ d_1 &&
r_2 \atsign{} c_2 \wedge c_2' \simparrow g_2 ~|~ d_2 
\end{eqnarray*}
is the pair of states 
$\tuple{c_1 \wedge d_2, True, g_1 \wedge g_2 \wedge c_1' = c_2', \emptyset}$ and
$\tuple{c_2 \wedge d_1, True, g_1 \wedge g_2 \wedge c_1' = c_2', \emptyset}$
where $g_1 \wedge g_2 \wedge c_1' = c_2'$ is satisfiable.

Deciding confluence requires that the 
CHR program is terminating. 
There are a number of syntactic 
restrictions on Haskell class and instance declarations
that will assure us that the resulting CHR programs are terminating.
There are also a number of other approaches to proving termination of 
CHR programs~\cite{fruehwirth:habilitacion}.

\section{Type Inference} \label{sec:inference}

Haskell is an implicitly typed language.  The task of type inference is to
infer a type for a given program or report error if the program is not
typable.  We identify the following three issues:
\begin{enumerate}
 \item Class and instance declarations must be correct.
       
 \item Type inference must generate a correct set of constraints which represent the possible solutions
       to the typing problem.
       The program is typable if the constraint problem is solvable.   
 \item Simplification of constraints is important for two reasons.
        Syntactically, it allows us to present type class constraints to the programmer in a more 
        readable form.
        Operationally, simplification allows us to put type class constraints into a 
        more efficient form. Type class constraints are translated into dictionaries. Hence, 
        simplifying type class constraints may allow a more efficient translation.
        This form of simplification is known as context reduction in Haskell.
\end{enumerate}

Type inference starts by processing all class and instance declarations,
i.e. we translate all class and instance declarations into a CHR program.
Then, type inference generates the constraints of the Haskell program and
applies the CHR solving process. In CHR, a constraint~$C$ is solvable if the derivation
from~$C$ does not lead to a constraint including~{\it False}.
Simplification may be invoked if necessary.
The following three sections expand on the three issues.

\subsection{Class and instance declarations}

In this section we show how to translate class and instance
definitions into CHRs.

\subsubsection*{Class definitions}

A class definition 
$$
\mclass~(d_1,\ldots,d_m) \Rightarrow \mClass~ x_1~ \ldots~ x_n ~\mwhere~ \ldots
$$ 
constrains any instance of the class \texttt{C} to 
also satisfy the class constraints $d_1,\ldots,d_m$.
Hence the corresponding CHR is
$$
\mClass~ x_1~\ldots~ x_n \proparrow d_1,\ldots,d_m
$$

\begin{example}
Consider the standard prelude definitions of @Ord@ and its translation:
$$\begin{array}{ll}
\mbox{\tt class Eq t => Ord t where ...} & \mlabel{S1} \atsign{} Ord~ t \proparrow Eq~ t 
\end{array}
$$
Whenever we assert the $Ord~ t$ constraint we must also satisfy the
$Eq~t$ constraint.
\end{example}

\subsubsection*{Instance definitions}

An instance definition
$$
\minst~(d_1,\ldots,d_m) \Rightarrow \mClass~ t_1~ \ldots~ t_n ~\mwhere~ \ldots
$$
maintains that tuple $t_1~ \ldots~ t_n$ is an instance of $\mClass$
if the constraints $d_1,\ldots,d_m$ are also satisfied. 
This corresponds to a simplification rule.
$$
\mClass~ t_1~\ldots~ t_n \simparrow d_1,\ldots,d_m
$$

\begin{example}
The instances of $Ord$ and $Eq$ for Lists and their translations are:
$$\begin{array}{ll}
\mbox{\tt instance Eq t => Eq [t] where ...}   & \mlabel{S2} \atsign{} Eq~ [t] \simparrow Eq~ t        \\
\mbox{\tt instance Ord t => Ord [t] where ...}  & \mlabel{S3} \atsign{} Ord~ [t] \simparrow Ord~ t   
\end{array}
$$
This means that we can prove that a type $[t]$ is an instance of
the class $Eq$ or $Ord$ if and only if we can prove that $t$ is an instance.
\end{example}

\subsubsection*{Checking instance definitions}

An instance declaration must be compatible with the class definition.
An instance declaration is correct 
in this sense if the resulting CHR program is
confluent. 

\begin{example}
If the instance declaration for @Ord [t]@ didn't require that @t@ was
also in class @Ord@, i.e.
$$\begin{array}{ll} 
\mbox{\tt instance Ord [t] where ...} & \mlabel{S4}  \atsign{} Ord~ [t] \simparrow True 
\end{array}
$$
We would have a non-confluent CHR program, since @Ord [t]@ has two
derivations whose result is not joinable:
$$ 
\begin{array}{rll}
&& \tuple{Ord~[t], True, True, \emptyset}_{\{t\}} \\
&\rightarrowtail& \tuple{True, True, Ord~[t],
\{\mlabel{S1}\atsign{}Ord~[t]\}}_{\{t\}} \\
\mlabel{S1}\atsign{} ~~Ord~[t] \proparrow Eq~[t] &\rightarrowtail&
\tuple{Eq~[t], True, Ord~[t], \emptyset}_{\{t\}} \\
&\rightarrowtail& \tuple{True, True, Ord~[t] \wedge
Eq~[t],\emptyset}_{\{t\}} \\
\mlabel{S4}\atsign{}~~Ord~[t] \simparrow True 
&\rightarrowtail& \tuple{True, True, 
Eq~[t],\emptyset}_{\{t\}} \\
\mlabel{S2}\atsign{}~~Eq~[t] \simparrow Eq~t 
&\rightarrowtail& \tuple{Eq~t, True, 
True,\emptyset}_{\{t\}} \\
&\rightarrowtail& \tuple{True, True, 
Eq~t,\emptyset}_{\{t\}} 
\end{array}
$$
and
$$ 
\begin{array}{rll}
&& \tuple{Ord~[t], True, True, \emptyset}_{\{t\}} \\
&\rightarrowtail& \tuple{True, True, Ord~[t],
\{\mlabel{S1}\atsign{}Ord~[t]\}}_{\{t\}} \\
\mlabel{S4}\atsign{}~~Ord~[t] \simparrow True 
&\rightarrowtail& \tuple{True, True, 
True,\emptyset}_{\{t\}} \\
\end{array}
$$
\end{example}

After generating a CHR program we can (assuming that its terminating)
check that it is confluent.  If it isn't confluent then there is an
error in the instance declarations, otherwise we can safely use the
CHR program for type inference.

\subsection{Solving Type Class Constraints}

With this view of type classes simply as constraints defined by CHR
rules, the solving process is obvious: generate the constraints of the
program text and apply the CHR solving process.

\begin{example} \label{ex:f}
Consider the Haskell function
\begin{quote}
\begin{code}
f g h = c where a = tail g; b = init h; c = a < b
\end{code}
\end{quote}
the constraints generated are 
\begin{equation} \label{ex:one}
 \ba{c}
  t_a = [t_1] \wedge t_g = [t_1] \wedge t_b = [t_2] \wedge t_h = [t_2] \wedge 
  t_f = t_g \arrow t_h \arrow t_c ~\wedge \\
  Ord~t_3 \wedge t_3 \arrow t_3 \arrow Bool = t_a \arrow t_b
  \arrow t_c
 \ea
\end{equation}
Note that we use subscript notation to associate expressions and 
their inferred type.
After simplification through unification we obtain
\begin{equation} \label{ex:two}
 \ba{c}
    t_a = t_g = t_b = t_h = t_3 = [t_1] \wedge t_2 = t_1 \wedge 
    t_f = [t_1] \arrow [t_1] \arrow Bool \wedge 
    Ord~[t_1]
 \ea
\end{equation}
If we now apply the constraint handling rules above to the
constraint $Ord~[t_1]$ we obtain
the following derivation:
$$
\begin{array}{rll}
&& \tuple{Ord~[t_1], True, True, \emptyset}_{\{t_1\}}  \\
& \rightarrowtail & \tuple{True, Ord~[t_1], True, \{\mlabel{S1}\atsign{}Ord~[t_1]\}}_{\{t_1\}} \\
~~\mlabel{S3} \atsign{} ~~Ord~[t_1] \simparrow Ord~t_1  & \rightarrowtail & \tuple{Ord~t_1, True,
True, \emptyset}_{\{t_1\}} \\
 & \rightarrowtail & \tuple{True, Ord~t_1, True, \{\mlabel{S1}\atsign{}Ord~t_1\}}_{\{t_1\}} \\
~~\mlabel{S1} \atsign{} ~~Ord~t_1 \proparrow Eq~t_1  & \rightarrowtail & 
   \tuple{Eq~t_1, Ord~t_1, True, \emptyset }_{\{t_1\}} \\
& \rightarrowtail^* & \tuple{True, Eq~t_1 \wedge Ord~t_1, True, \emptyset}_{\{t_1\}} 
\end{array}
$$
Since such derivations are laborious, from now on we will 
use simplified derivations where the tuple is represented as a single
conjunction and unification is applied to remove extra variables,
and the token set is omitted.
The corresponding derivation is
$$
Ord~[t_1]\rightarrowtail_{S3} Ord~t_1 \rightarrowtail_{S1} Ord~t_1  \wedge
Eq~t_1
$$
Since the CHR program is confluent,  all alternative rewritings
must (eventually) give the same result, for example
$$
Ord~[t_1] \rightarrowtail_{S1} Ord~[t_1] \wedge Eq~[t_1]
\rightarrowtail_{S2} Ord~[t_1]  \wedge Eq~t_1 \rightarrowtail_{S3}
Ord~t_1  \wedge Eq~t_1  
$$
gives the same answer.
\end{example}

\subsection{Presenting Type Class Constraints}

CHRs are also simplification rules, they replace
a (type class) constraint by a simpler, equivalent form.
When we want to present a type definition to the user we wish to have the
``simplest'' possible form.  This is contrary to the usual solving methodology that
adds redundant information to simplify the detection of unsatisfiability.
To this end it is worth adding a separate simplification phase
for presenting constrained types to users.
This too can be represented by a (disjoint) CHR program.

The presentation rules for class definition
$$
\mclass~(d_1,\ldots,d_m) \Rightarrow \mClass~ x_1~ \ldots~ x_n ~\mwhere~ \ldots
$$ 
are of the form
$\mClass~ x_1~\ldots~ x_n,~d_i \simparrow \mClass~ x_1~\ldots~ x_n$
which removes the redundant constraints for presentation.

\begin{example}
An example class definition and its corresponding presentation rule are:
$$\begin{array}{ll}
\mbox{\tt class Eq t => Ord t where ...}  & \mlabel{P1} \atsign{} Ord~ t, Eq~t \simparrow Ord~ t  
\end{array}
$$
In presenting the answer to Example~\ref{ex:f} we obtain the
type @f :: Ord t => t -> t -> Bool@ since
$ Ord~t_1, Eq~t_1 \rightarrowtail_{P1} Ord~t_1$.
\end{example}

\section{Properties of Type Class Systems} \label{sec:properties}

CHRs allow us to characterize under which conditions we 
retain feasible type inference
and unambiguous programs.

\subsection{Feasible Type Inference}  \label{sec:inf-properties}

Feasible type inference must be decidable and yield principal types.
In CHR, type inference is decidable if the CHR program is terminating.
If we restrict instance and class definitions to those allowed by the
Haskell report and GHC's multi-parameter type class extensions then
the resulting CHR program is always terminating.

Principality means that for a given Haskell program the inferred type subsumes
all other types we could possibly give to this program.
CHRs preserve principal types because they only ever map constraints to logically
equivalent constraints.
Note that principal types are
not syntactically unique (in Example~\ref{ex:f}, (\ref{ex:one}) and (\ref{ex:two})
are both possible principal types) but since the CHR program is
confluent, we will always present the same type.

Confluence of the CHR program is a vital property. It guarantees that
the CHR program is consistent, so we can meaningfully talk about
unsatisfiable type constraints, and it guarantees that instance
definitions satisfy class definitions. 

\subsubsection*{Type Signatures}

Type signatures allow the user to declare that variables have a certain type.
They are an optional form of program documentation but necessary, for example, 
to retain decidability in the case of polymorphic recursion.
In the presence of type signatures, we are moving from a type inference
problem to a type reconstruction problem.
This shift implies that our constraint solver also needs to handle entailment among constraints.
Decidable constraint entailment is often difficult to establish.
As in~\cite{jones:typing-haskell-in-haskell},
we find that the conditions of Haskell 98 are sufficient to 
guarantee that entailment is decidable.

\subsection{Unambiguous Types} \label{sec:ambiguity}

In the Haskell framework each function must have an unambiguous type. 
We can understand the notion of \emph{unambiguity} in terms of CHRs,
so that later when we extend the type system to use more 
complex CHRs we retain this property.

Suppose the inferred type for function $f$ is
$f :: D \Rightarrow \tau$
then the type for $f$ is \emph{unambiguous} for CHR program $P$ if, 
for renaming to new variables $\rho$,
$P \models 
D \wedge \rho(D) \wedge \tau = \rho(\tau) \rightarrow \alpha =
\rho(\alpha)$
for each variable $\alpha \in vars(D \Rightarrow \tau)$.

We have a sound check for the unambiguity of a type using the CHR
program $P$ by seeing if $D \wedge \rho(D) \wedge \tau = \rho(\tau)
\rightarrowtail_P C$ where $\models C \rightarrow \alpha =
\rho(\alpha)$.  This check is complete for Haskell 98 programs,
ignoring Haskell's Numeric defaulting mechanism.  We conjecture that
it is complete for other interesting Haskell extensions, such as
Functional Dependencies.

\begin{example}\label{ex:collect}
Recall the @Collects@ example from Section~\ref{sec:motivation}.
The type of @empty@ is
$empty :: Collects~ e~ ce \Rightarrow ce$.
This type is ambiguous, because
$$Collects~ e~ ce \wedge Collects ~e' ~ce' \wedge ce = ce'$$
does not allow any propagation steps, and does not imply that
$e = e'$. 

The type defined for @insert@ is
$insert :: Collects~ e~ ce \Rightarrow e \arrow ce \arrow ce$
which is clearly unambiguous.
\end{example}

\section{Understanding Type Classes Extensions} \label{sec:extensions}

There are a number of extensions to type classes that have been proposed in the literature.
These can
be understood in the uniform framework of CHRs.  By unifying the
representation of different extensions we can gain insight into what kinds
of extensions are feasible.

\subsubsection*{Functional Dependencies}

Jones~\cite{JonesESOP2000} proposes an extension of multi-parameter
type classes to include functional dependencies among class arguments. 
From a CHR point of view, 
functional dependencies among variables in a type class
just extend the proof requirements for an instance.
They are expressible straightforwardly using CHRs.

A class definition with functional dependencies has the form
$$
\mclass~(d_1,\ldots,d_m) \Rightarrow \mClass~ x_1~ \ldots~ x_n ~|~ \mbox{\it fd}_1, \ldots, \mbox{\it fd}_k
 ~\mwhere~ \ldots
$$ 
where $\mbox{\it fd}_i$ is a \emph{functional dependency} of the form
$(x_{i_1},\ldots,x_{i_k}) \rightsquigarrow
x_{i_0}$.
In~\cite{JonesESOP2000} the rhs of the $\rightsquigarrow$ can
have a list of variables. We use this simpler form, the expressiveness is
equivalent.
The functional dependency asserts that given fixed values of 
$x_{i_1},\ldots,x_{i_k}$ then there is only one value of
$x_{i_0}$ for which the class constraint $\mClass~ x_1~ \ldots~ x_n$
can hold.

The CHR translation creates a propagation rule for each 
functional dependency of the form
$$
\mClass~ x_1 ~\ldots~ x_n, \mClass~ y_1 ~\ldots~ y_n \proparrow 
x_{i_1} = y_{i_1} \wedge \cdots \wedge x_{i_k} = y_{i_k} ~|~ x_{i_0} = y_{i_0} 
$$
This CHR enforces the functional dependency.

\begin{example} \label{ex:collect2}
Returning to the collection class example, but
now adding a functional dependency.
We have the following rule and (simplified) CHR:
$$\begin{array}{ll}
\mclass ~Collects ~e ~ce ~|~ ce \rightsquigarrow e ~\mwhere~ ... 
& \mlabel{T1} \atsign{} Collects ~e ~ce, Collects ~f ~ce \proparrow f = e 
\end{array}
$$

Note now that the type for @empty@ is unambiguous because
\begin{eqnarray*}
 Collects~ e~ ce \wedge Collects ~e' ~ce' \wedge ce = ce' &
 \rightarrowtail_{T1} & Collects~ e~ ce \wedge Collects ~e' ~ce' \wedge ce =
ce'\wedge e = e'
\end{eqnarray*}
\end{example}

\begin{example}
The type checking/inference for
\begin{quote}
\begin{code}
f x y c = insert x z where z = insert y c
\end{code}
\end{quote}
where @insert :: Collects e ce => e -> ce -> ce@ gives
\begin{eqnarray*}
&& Collects~e~ce \wedge e \arrow ce \arrow
ce = t_y \arrow t_c \arrow t_z \wedge Collects~ e'~ce' \wedge 
e' \arrow ce' \arrow
ce' = t_x \arrow t_z \arrow r   \\
& \equiv & Collects~e~ce \wedge Collects~ e'~ce \wedge 
t_c = t_z = ce' = r = ce \wedge t_y = e \wedge t_x = e' \\
& \rightarrowtail_{T1} & Collects~e~ce \wedge Collects~ e'~ce \wedge 
t_c = t_z = ce' = r = ce \wedge t_y = e \wedge t_x = e' \wedge e = e' \\
 & \equiv & Collects~e~ce \wedge t_c = t_z = ce' = r = ce \wedge t_x = t_y = e
\end{eqnarray*}
The type inferred is 
@f :: Collects e ce => e -> e -> ce -> ce@ as expected.
\end{example}

This view of functional dependencies as CHRs clarifies one
of the questions that Jones poses in the end of~\cite{JonesESOP2000}.
Given the declarations
\begin{eqnarray*}
&& \mclass~U~ a~ b ~|~ a \rightsquigarrow b~ \mwhere ... \\ 
&& \mclass~U~ a~ b \Rightarrow V~ a~ b~ \mwhere ...  
\end{eqnarray*}
in Jones' framework, from the constraints
$U~ a~ b \wedge V~ a~c$ it cannot be inferred that $b = c$. 
The CHR rules support the automatic inference of inherited functional dependencies
Consider the following example:
$$
U~ a~ b \wedge V~ a~ c ~~\rightarrowtail~~ U~ a~ b \wedge V~ a~ c \wedge U~ a~ c
~~\rightarrowtail~~ U~ a~ c \wedge V~ a~ c \wedge b = c
$$

\subsubsection*{Constructor Classes}

Type constructors are simply a functional relation among
types. We can understand them simply using CHRs, this is
simply a matter of replacing constructor expressions 
$f~e$ (lets say $\equiv \mbox{\it fe}$)
by explicit kind constraints 
$\kindone{}~ f~ e~ \mbox{\it fe}$.
The class constraints need to satisfy appropriate properties
(functionality, surjectiveness) which can be expressed with CHRs,
as well as the kind constraints. For example

\begin{eqnarray*}
\mbox{functional} &\atsign& \kindone~ f~ e~ \mbox{\it fe}, ~\kindone~ f~ e~ \mbox{\it fe}' ~~\proparrow~~ \mbox{\it fe} = \mbox{\it fe}' \\
\mbox{surjective} &\atsign& \kindone~ f~ e~ \mbox{\it fe}, ~\kindone~ f'~ e'~ \mbox{\it fe} ~~\proparrow~~ f =
f', ~e = e'\\
\mbox{kinding} &\atsign& \kindone~ f~ e~ \mbox{\it fe}, ~\kindzero~ f ~~\simparrow~~ False 
\end{eqnarray*}

Clearly constructor classes can be expressed using CHRs, and hence we
have a more uniform understanding of their meaning and use.
The presentation of constructor class constraints
to the user might be preferable with the usual notation,
but this is simply a matter of presentation.

\section{Further Extensions to Types Classes Using CHRs} 
\label{sec:new-extensions}

Given we can use CHRs to specify existing 
type class extensions, an immediate
question is what other new extensions can we express in
terms of CHRs. 

\subsubsection*{Disjointness of Type Classes}

The example in Section~\ref{sec:motivation} illustrates how it
may be useful to have additional constraints on the instances of a
class.  With this disjointness we may be able to have a weaker
definition of non-overlapping instances.

\begin{example}
A CHR expressing that the @Integral@ and @Fractional@
type classes are disjoint is simply.
\begin{eqnarray*}
\mlabel{IF} \atsign{}  Integral~ t, Fractional~t \simparrow False 
\end{eqnarray*}
If we translate the two instance declarations for
$Dividable$ from the motivation we obtain:
$$
\begin{array}{lll}
& \mlabel{DI1} \atsign{} Dividable~t \simparrow Integral~t  \\
& \mlabel{DF1} \atsign{} Dividable~t \simparrow Fractional~t 
\end{array}
$$
Clearly the resulting CHRs are not confluent, since
there are two disjoint replacements for $Dividable~t$.
But we could weaken the simplification rules to
$$\begin{array}{lll}
& \mlabel{DI2} \atsign{} Dividable~t, Integral~t \simparrow Integral~t  \\
& \mlabel{DF2} \atsign{} Dividable~t, Fractional~t \simparrow Fractional~t  
\end{array}
$$
which together with $(IF)$ give a confluent system. Note that
with this reading, we can remove a $Dividable~t$ constraint
if we already know that $t$ is in $Integral$ or $Fractional$
but we cannot simply replace the $Dividable~t$ with one of
these constraints.
\end{example}

Another extension is to allow negative information about type
classes. 

\begin{example}
The intention of the @Num@ class is to describe numeric types.
We might insist that
functional types are never numbers
by adding the rule
$$
\mlabel{N1} \atsign{} \mbox{\it Num}~ (s \arrow t) \simparrow False
$$
Or we might form a @NotNum@ class meant to indicate types which cannot be
numbers, where functional types are in this class, expressed by the rules
\begin{eqnarray*}
&& \mlabel{N2} \atsign{} \mbox{\it NotNum}~ t, \mbox{\it Num}~ t \simparrow False \\
&& \mlabel{N3} \atsign{} \mbox{\it NotNum}~ (s \arrow t) \simparrow True
\end{eqnarray*}
Then, in either case, the declaration @ instance Num (a->b) where ... @
will cause an error to be detected since the resulting CHR program is
not confluent.
\end{example}

In general there are considerable problems supporting overlapping
class instances. 
The key thing to understand
is that \emph{confluence} of the resulting CHRs gives the 
behaviour we
expect. If the resulting CHR program is not confluent, then there is
an error with the program's class and instance definitions. 
If the CHR program is confluent it doesn't mean there are not problems,
but at least the correctness of types is not affected by the
overlapping instances.

\begin{example}
Consider the program
\begin{quote}
\begin{code}
instance A t => C Bool t where ... 
instance B s => C s Int  where ...
\end{code}
\end{quote}
The question is what should happen for class constraint $C~ Bool~ Int$.
If @Bool@ is in $B$ and @Int@ is in $A$, then it is clear that the
constraint holds, we must simply choose which instance's methods to use.
If we allow overlapping class instances, this leaves the fundamental problem of how to make this choice. 
\end{example}

\subsubsection*{Record Types}

We can define an extensible record type class by using a set of
types for labels, and two class constraints:
\begin{quote}
\begin{code}
class Rec r l b where
    select :: r -> l -> b          -- access record r, b = r.l
    update :: r -> l -> b -> r     -- update record r so r.l = b
class Rec r2 l b => Ext r1 l b r2 where
    extend :: r1 -> l -> b -> r2
\end{code}
\end{quote}
The $Rec$ constraint constrains $r$ to be a record type
containing element labeled $l$ of type $b$.
The $Ext$ constraint constraints $r2$ to be a type obtained
by extending $r1$ with a new element labelled $l$ of type $b$.
There are some obvious rules we want to hold in order to enforce
type correctness.
\begin{eqnarray*}
\mbox{class-defn} &\atsign& Ext ~r_1~l~b~r_2 \proparrow  Rec~r_2~l~b \\
\mbox{functionality} &\atsign& Rec ~r~ l~ b_1, Rec~r~l~b_2 \proparrow b_1 = b_2 \\
\mbox{false-extension} &\atsign& Ext ~r_1~l~b_1~r_2, Rec~r_1~l~b_2 \simparrow False \\
\mbox{extension} &\atsign& Ext ~r_1~l_1~b_1~r_2, Rec~r_2~l_2~b_2 \proparrow l_1 \neq l_2 ~|~ Rec~r_1~l_2~b_2 
\end{eqnarray*}
Consider the code
\begin{quote}
\begin{code}
f x = (select x (A), select x (B)) -- (X) is unique element of type X
g x y l = extend y l (select x l)
h x y = if x == y  then [select x (A), select y (B)] else []
\end{code}
\end{quote}
The (simplified) type constraints for $f$, $g$ and $h$ are 
\begin{eqnarray*}
&& f :: (Rec ~t_x~ A~ t_{f_1}, Rec~ t_x~ B~ t_{f_2}) \Rightarrow t_x \arrow
(t_{f_1}, t_{f_2}) \\
&& g :: (Rec ~t_x~ t_l~ t_s, Ext~ t_y~ t_l~ t_s~t_e) \Rightarrow t_x \arrow t_y
\arrow t_l \arrow t_e  \\
&& h :: (Rec ~t_x~ A~ t_e, Rec~ t_x~ B~ t_e) \Rightarrow t_x \arrow t_x \arrow [t_e]
\end{eqnarray*}

\section{Conclusion} \label{sec:conclusion}

We have demonstrated that constraint handling rules are a useful tool in
understanding type class systems.  Several existing type class systems can
be expressed in terms of CHR rules.  Surprisingly, constructor
classes and multi-parameter classes which have been considered to be
orthogonal extensions are both expressible in terms of CHRs.  Other useful
extensions such as disjoint classes can also naturally be expressed in CHRs.
Feasible type inference and unambiguity are important issues in the design
of a type class system.  CHRs allow us to characterize sufficient
conditions under which we can retain both properties.  
We conclude that CHRs offer a natural way to study type class systems.
In this work, the development was rather motivated by examples and intuition.
We are currently working on a more 
formal treatment which we will report separately.

\small

\newcommand{\etalchar}[1]{$^{#1}$}

\end{document}